%% file: sensing_arxiv_v2.tex
\documentclass[aps,pra,twocolumn,superscriptaddress,floatfix]{revtex4-1}
\usepackage{graphicx}
\usepackage{amsmath}
\usepackage{amssymb}
\usepackage{mathrsfs}
\usepackage{amsfonts}
\usepackage{dsfont}
\usepackage{dcolumn}
\usepackage{bm}
\usepackage{booktabs}
\usepackage{tabularx}
\usepackage{hyperref}
\usepackage{textcomp}
\usepackage{multirow}
\usepackage{siunitx}
\usepackage[rgb]{xcolor}
\usepackage{pgf}
\usepackage[abs]{overpic}
\usepackage{calc}
\usepackage{pstricks}
\usepackage{capt-of}

\renewcommand{\Re}{{\rm Re}}

\newcommand{\dd}{\mathrm{d}}

\newcommand{\eref}[1]{Eq.~(\ref{#1})}

\newcommand{\br}{\boldsymbol{r}}

\newcommand{\bE}{\boldsymbol{E}}

\newcommand{\etal}{\emph{et al}.~}

\DeclareRobustCommand{\AffOne}{%
    Max Planck Institute for the Science of Light, 
    G\"{u}nther-Scharowsky-Stra{\ss}e 1/Bldg.~24, 
    91058 Erlangen, 
    Germany
}
\DeclareRobustCommand{\AffTwo}{%
    Institute for Optics, Information and Photonics, 
    Universit\"{a}t~Erlangen-N\"{u}rnberg, 
    Staudtstra{\ss}e 7/B2, 
    91058 Erlangen, 
    Germany
}
\DeclareRobustCommand{\AffThree}{%
    Erlangen Graduate School in Advanced Optical Technologies (SAOT), 
    Paul-Gordan-Stra{\ss}e 6, 
    91052 Erlangen, 
    Germany
}
\DeclareRobustCommand{\AffFour}{%
    Department of Physics, University of Ottawa, 
    25 Templeton, Ottawa, Ontario, 
    K1N 6N5 Canada
}
\DeclareRobustCommand{\AffFive}{%
    Institute of Applied Physics, 
    Friedrich-Schiller University Jena, 
    Max-Wien Platz 1, 
    07743 Jena, Germany
}
\DeclareRobustCommand{\AffSix}{%
    Laboratoire Kastler Brossel, 
    Universit\'e Pierre~et~Marie Curie, 
    \'Ecole~Normale~Sup\'erieure, 
    CNRS, 
    4 place Jussieu, 
    75252 Paris, France
}

\begin{document}
\title{Classically entangled optical beams for high-speed kinematic sensing}
\author{Stefan Berg-Johansen}
\affiliation{\AffOne}
\affiliation{\AffTwo}
\author{Falk T\"oppel}
\affiliation{\AffOne}
\affiliation{\AffTwo}
\affiliation{\AffThree}
\author{Birgit Stiller}
\affiliation{\AffOne}
\affiliation{\AffTwo}
\author{Peter Banzer}
\affiliation{\AffOne}
\affiliation{\AffTwo}
\affiliation{\AffFour}
\author{Marco Ornigotti}
\affiliation{\AffFive}
\author{Elisabeth Giacobino}
\affiliation{\AffOne}
\affiliation{\AffSix}
\author{Gerd Leuchs}
\affiliation{\AffOne}
\affiliation{\AffTwo}
\affiliation{\AffFour}
\author{Andrea Aiello}
\affiliation{\AffOne}
\affiliation{\AffTwo}
\author{Christoph Marquardt}
\affiliation{\AffOne}
\affiliation{\AffTwo}
\date{\today}
\maketitle
%
%
%
\noindent\textbf{%
Tracking the kinematics of fast-moving objects is an important diagnostic tool
for science and engineering. Existing optical methods include 
high-speed CCD/CMOS imaging \cite{whybrew-high-speed-2004},
streak cameras \cite{velten-picosecond-2011}, 
lidar \cite{weitkamp-lidar-2005},
serial time-encoded imaging \cite{goda-serial-2009}
and sequentially timed all-optical mapping \cite{nakagawa-sequentially-2014}.
Here, we demonstrate an entirely new approach to positional and directional
sensing based on the concept of classical entanglement
\cite{spreeuw-classical-1998,spreeuw-classical-2001,luis-coherence-2009}
in vector beams of light.
The measurement principle relies on the intrinsic correlations existing in such
beams between transverse spatial modes and polarization. The latter can
be determined from intensity measurements with only a few fast photodiodes,
greatly outperforming the bandwidth of current CCD/CMOS devices. In this way,
our setup enables two-dimensional real-time sensing with temporal resolution in
the GHz range. We expect the concept to open up new directions in
metrology and sensing.
}

%
%
%
%
Vector beams of light with cylindrical, non-uniform polarization patterns 
\cite{zhan-cylindrical-2009} 
have found application in diverse areas of optics such as 
improved focusing 
\cite{dorn-sharper-2003}, 
laser machining \cite{meier-material-2007}, 
plasmon excitation \cite{mojarad-tailoring-2009}, 
metrology \cite{fatemi-cylindrical-2011}, 
optical trapping \cite{kozawa-optical-2010}
and nano-optics 
\cite{%
kindler-waveguide-2007%
,neugebauer-polarization-2014%
,wozniak-selective-2015}.
Recently, vector beams have attracted attention
\cite{%
holleczek-poincare-2010,
borges-bell-like-2010,
karimi-spin-orbit-2010,
qian-entanglement-2011,
kagalwala-bells-2012}
due to a simple but striking property: when viewed as a superposition of
transverse electromagnetic modes with orthogonal linear polarizations, the
nonseparable mode function of a radially polarized vector beam is
mathematically equivalent to a maximally entangled Bell state of two qubits
known from quantum mechanics.
In contrast with the canonical Bell states in quantum optics, where two photons
are entangled in polarization and exhibit non-local correlations when spatially
separated, this ``classical entanglement'' in vector beams is necessarily local
as it exists only between different degrees of freedom of one and the same
physical system 
\cite{aiello-classical-2014}. 

However, these correlations have recently been shown to represent a valuable
resource. Vector beams have been used to violate an analogue of Bell's
inequality for spin-orbit modes
\cite{borges-bell-like-2010,karimi-spin-orbit-2010} 
and have led to continuous-variable entanglement between different degrees of
freedom 
\cite{gabriel-entangling-2011}. 
In addition, vector beams have been used to implement classical counterparts of
quantum protocols 
\cite{%
oliveira-implementing-2005,
hashemi-rafsanjani-teleportation-2015}. 
Promising proposals include an application to the study of quantum
random walks 
\cite{goyal-implementing-2013} 
and real-time single-shot Mueller matrix measurements 
\cite{toppel-classical-2014}, and a scheme for measuring the depolarization
strength of a material has been implemented \cite{fade-depolarization-2012}.
In the present work, we demonstrate for the first time a fully operational
application of classical entanglement to high-speed kinematic sensing.

%
\begin{figure}[t]
    \def\svgwidth{0.45\textwidth} 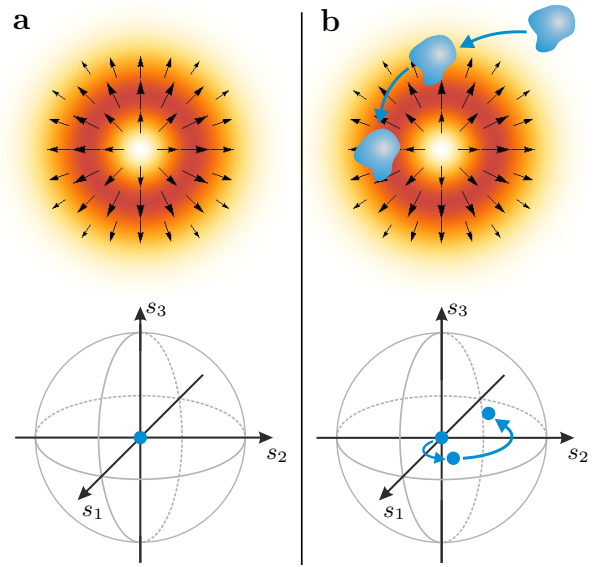
    \caption{%
        \textbf{Classical entanglement.}
        \textbf{a.}~\textit{(top)} The transverse electric field distribution
        of a radially polarized beam of light. The orange density plot shows
        the beam's doughnut-shaped intensity distribution, while black arrows
        indicate the position-dependent instantaneous local direction of the
        electric field vector.
        \mbox{\textit{(bottom)}}~The beam's global polarization state is shown
        in a Poincar\'e sphere representation. Initially, the light field is
        globally unpolarized, with the Poincar\'e vector located at the origin.
        \textbf{b.}~If an opaque obstacle is brought into the beam, the global
        polarization takes on non-zero values according to the obstacle's position
        within the beam. This method allows the object's kinematics to be
        inferred from a polarization measurement alone.%
    }
    \label{fig:mode-theory}
\end{figure}
%

Several techniques are nowadays available for sensing the kinematics of
fast-moving objects
\cite{%
whybrew-high-speed-2004,
velten-picosecond-2011,
weitkamp-lidar-2005,
goda-serial-2009,
nakagawa-sequentially-2014}.
Each comes with its own strengths and drawbacks. For example, high-speed
imaging is typically limited to capturing only a small number of frames, while
pump-probe techniques require the recorded event to be repeated identically
many times.
Ideally, one seeks a solution that is capable of performing fast sensing
continuously, in real-time and from a simple setup, employing only standard
equipment and offering flexibility in the choice of wavelength.
By using the nonseparable mode structure of cylindrically polarized beams (see
Figure~\ref{fig:mode-theory}), one only needs to detect changes in
polarization, thus fulfilling all the above requirements at the same time.
In the following, we first discuss the physics of vector beams and then
introduce the technique of sensing and show the results of our experimental
investigations.

The electric field of a general non-uniformly polarized paraxial beam can be
written as:
\begin{align}\label{radial}
\bm{E}(\bm{\rho},z) =   \bm{e}_1  f_1(\bm{\rho},z) + \bm{e}_2 \,  f_2(\bm{\rho},z),
\end{align}
where the vectors $\bm{e}_1, \, \bm{e}_2 $  determine the beam polarization, 
the scalar functions $f_1(\bm{\rho},z), \,f_2(\bm{\rho},z)$ set the wavefront
and 
$\bm{\rho} = \hat{\bm{x}} x +\hat{\bm{y}} y$ 
is the transverse position vector (see ``Methods'' below). 
The expression \eqref{radial} is \emph{nonseparable}, namely it is not possible
to rewrite it as the simple product of only one polarization vector and
a single scalar function. In this sense, Eq.~\eqref{radial} has the same
mathematical structure as a two-qubit entangled quantum state
\cite{aiello-classical-2014}.
It is a well-established result of mathematical physics that any
two-dimensional field of the form \eqref{radial} can be recast in the so-called
Schmidt form 
${\bm{E}(\bm{\rho},z) = 
\sqrt{\lambda_1} \,\hat{\bm{u}}_1 v_1(\bm{\rho},z) 
+  \sqrt{\lambda_2} \,\hat{\bm{u}}_2 v_2(\bm{\rho},z)}$ 
where 
$\{\hat{\bm{u}}_1 ,\hat{\bm{u}}_2  \}$ 
and
$\{v_1,v_2 \}$ 
form complete orthonormal bases in the polarization and spatial mode vector
spaces, respectively, with 
$\lambda_1 \geq \lambda_2 \geq 0$. 
If either 
$\lambda_1 = 0$ 
or 
$\lambda_2 = 0$,  
the expression of
$\bm{E}(\bm{\rho},z)$ 
is factorable and the beam is uniformly polarized. Vice versa, if 
$\lambda_1 \lambda_2 \neq 0$,  
the beam displays a non-uniform polarization pattern and is said to be
``classically entangled''. 
Thus, analogously to a bona fide quantum state, in a non-uniformly polarized
beam, polarization and spatial degrees of freedom are so strongly correlated
that if, by any means, one alters the beam's spatial profile, then the
polarization changes accordingly. Our sensing technique relies precisely upon
this peculiar phenomenon.
Owing to the classical entanglement exhibited by the beam, we are able
to retrieve information about the position of a moving object partially
obstructing the beam only by measuring the polarization of the
latter: no spatially resolving measurements are needed. 
Since polarization measurements can be performed at GHz rates, with our system
we are able to track very fast objects. 

For a field $\bm{E}(\bm{\rho},z)$ in the Schmidt form, the measurable Stokes
parameters can be written as
%
%
%
%
\begin{subequations}
    \label{stokes_parameters_optical}
    \begin{align}
    s_0 &= \lambda_1 + \lambda_2, \\
    s_1 &= \bigl(\lambda_1 - \lambda_2 \bigr)\bigl( \left|a_{x}\right|^2 -\left|a_{y}\right|^2 \bigr), \\
    s_2 &= \bigl(\lambda_1 - \lambda_2 \bigr)\bigl( a_{x} a_{y}^* + a_{x}^* a_{y} \bigr), \\
    s_3 &=  i \bigl(\lambda_1 - \lambda_2 \bigr)\bigl(a_{x} a_{y}^* - a_{x}^* a_{y} \bigr),
    \end{align}
\end{subequations}
where 
$a_x = \hat{\bm{u}}_1 \cdot \, \hat{\bm{x}}$ 
and 
$a_y = \hat{\bm{u}}_1 \cdot \, \hat{\bm{y}}$.
In a radially polarized beam one has 
$\lambda_1=\lambda_2$ and thus $s_1=s_2=s_3=0$,
reflecting the fact that such a beam appears completely unpolarized in the
absence of an obstruction.

When an \emph{opaque} object cuts across a non-uniformly polarized beam, the
spatial and polarization patterns of the latter vary with time according to the
obstructing object's instantaneous position, as described by its central
coordinates $x_0(t),y_0(t)$.
It is not difficult to show that for such a modified beam,
Eqs.~\eqref{stokes_parameters_optical} are still valid, provided that 
$\lambda_1, \lambda_2,a_x ,a_y $ 
are regarded now as functions of
$x_0(t),y_0(t)$. 
When the values of the Stokes parameters 
$s_0, s_1, s_2, s_3$ 
are replaced by the measured ones on the left side of Eqs.~\eqref{stokes_parameters_optical},
these can be regarded as a nonlinear algebraic system of four equations
for the two variables 
$x_0(t),y_0(t)$, 
which can be solved by means of suitable algorithms.
In this way, the instantaneous trajectory of the object is recovered.

%
\begin{figure*}[t]
    \centering
    {\def\svgwidth{0.80\textwidth}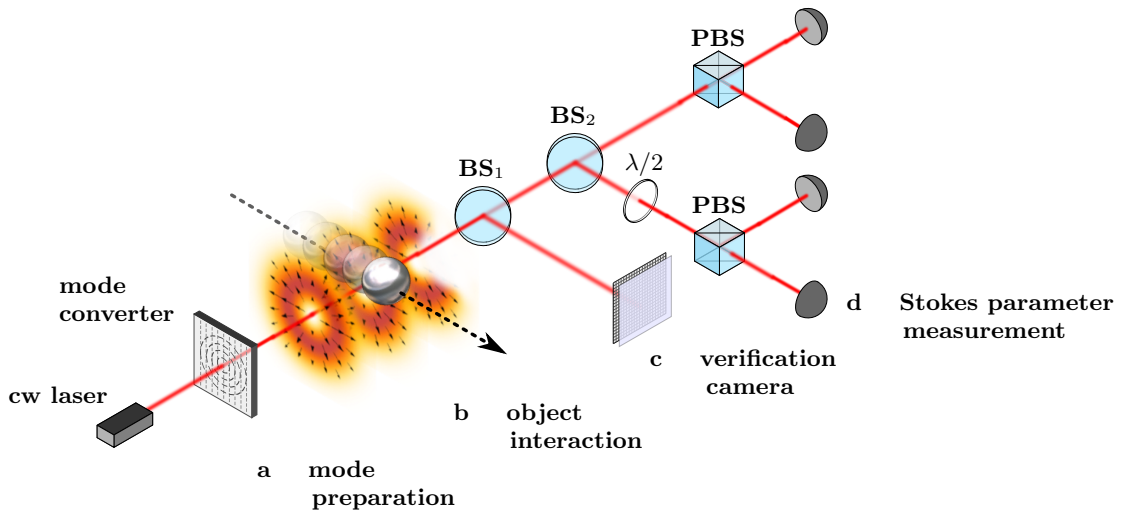}
    \caption{%
        \textbf{Experimental setup.}
        \textbf{a.}~A continuous-wave laser beam is prepared in a
        radially polarized mode by a liquid crystal mode converter.
        \textbf{b.}~The beam impinges on an opaque object whose
        motion in space modulates the beam's polarization Stokes parameters.
        \textbf{c.}~A polarization-independent beamsplitter (BS1)
        taps off \SI{10}{\percent} of the beam for inspection by a conventional
        camera. This allows for mode pattern characterization and
        independent verification measurements. 
        \textbf{d.}~A polarization-independent 50/50
        beamsplitter (BS2) divides the beam up for projection onto its
        linear polarization components via a pair of polarizing
        beamsplitters (PBS) and a half-wave plate ($\lambda/2$). The
        projections are simultaneously measured by four InGaAs detectors with
        \SI{4}{\giga\hertz} bandwidth. By linear combination of the projection
        signals, the beam's Stokes parameters are obtained.  Knowledge of the
        Stokes parameters allows the object's instantaneous trajectory to
        be reconstructed (see Figure~\ref{fig:results}).%
      }
    \label{fig:setup}
\end{figure*}
%
%

The experimental setup is shown in Figure~\ref{fig:setup}. We prepare
a continuous-wave laser beam in a radially polarized mode.
The beam impinges on a moving sample. Subsequently, half-waveplates and
polarizing beam splitters are used to project the beam onto its horizontal,
vertical, diagonal and anti-diagonal polarization components.
Finally, a network of four InGaAs photodetectors with \SI{4}{\giga\hertz}
\SI{3}{\decibel}-bandwidth measures the individual projections, from which the
Stokes parameters $s_0$, $s_1$ and $s_2$ can be straightforwardly obtained (see
``Methods''). For the particular case of a radially polarized mode, the $s_3$
parameter is always zero. An auxiliary camera is used for additional visual
verification and beam characterization.

In order to demonstrate the system's broad applicability, three types of
measurement are carried out.
First, a metal rotor is made to turn about the beam axis
(Figure~\ref{fig:results}a). By sampling the Stokes parameters during the
motion of the rotor, the instantaneous value of its angle of rotation
$\theta_0$ is succesfully inferred.  An accuracy of \SI{4.1}{\degree} (mean
error) is achieved without correcting for beam imperfections and detector
coupling. 

Second, a metal sphere is moved across the beam (Figure~\ref{fig:results}b). 
We measure the Stokes parameters with an acquisition time of
\SI{250}{\pico\second} at each position. A Bayesian algorithm is
used to estimate the sphere's position from these data (see Supplementary
Material).
The inferred trajectory shown in Figure~\ref{fig:results}b is seen
to be in good agreement with the actual trajectory.
As one expects, the inference is particularly successful in areas where the
beam has a high intensity, i.e.~where the Stokes parameter modulation
introduced by the sphere results in a higher signal-to-noise ratio.

Third and finally, the setup's real-time capability is demonstrated by
focusing the beam and measuring the Stokes parameters during the transit of
a knife edge moving at \SI{27(2)}{\metre\per\second} across the focal plane. 
(The beam is sufficiently gently focused that it is not dominated by
longitudinal field components at the waist, see ``Methods''.)
As seen from the captured data in Figure~\ref{fig:focusing}, the transit takes
only \SI{92}{\nano\second}, after which the beam is fully covered. From the
shape of the recorded traces, the knife edge's direction of motion,
horizontal in this case, can be inferred (up to a 180$^\circ$ rotation, see
Supplementary Material).
The event is captured as a sequence of single-shot measurements, requiring only
a single occurrence.  
Furthermore, since the measurement is triggered on a change in $s_0$, the
particular instant of occurrence does not have to be known in advance.
As the Stokes parameters are captured continuously there is no dead
time in this measurement.
This result clearly demonstrates the measurement technique's potential for
high-speed kinematic sensing. 
The technique allows for the use of bucket
detectors rather than spatial detectors, and the measurement can be as fast as
the detectors.
With the analog bandwidth of \SI{4}{\giga\hertz} available in our setup, we
would be capable of resolving even sub-nanosecond motions.
%

All three measurements confirm the setup's ability to perform quantitatively
meaningful kinematic sensing at very high temporal resolutions. 
We note that the measurement precision is subject to random error from the
electronic detector noise at high bandwidths. This becomes dominant in
the regime where the measured sample has only a small overlap with the beam (as
seen in Figure~\ref{fig:results}b), or when the sample covers the beam
completely. 
Some applications, such as precision sensing of objects moving within
a confined region, may benefit from using a beam with
a nonzero $s_3$ Stokes parameter. Such beams have been suggested for the
investigation of small particle scattering \cite{beckley-full-2010}. 
Although they require an additional photodiode pair, such beams avoid the
zero of intensity at the origin. 
We note, however, that the classical entanglement of such a beam is not
maximal, and that the correlations between polarization and position
are therefore necessarily weaker.

In summary, we have demonstrated that the classical entanglement manifested by
vector beams of light may be used to detect the kinematics of very fast
objects with \si{\giga\hertz} temporal bandwidth. The method is
possible because for cylindrically polarized beams, spatial information
may be acquired by measurements on the polarization degree of freedom, owing to
the classically entangled mode structure.
The method presented requires only standard optical components which are
commercially available at a wide range of optical wavelengths and can easily be
extended to the microwave regime. It allows for
continuous, real-time measurement of two-dimensional spatial information with
unprecedented temporal resolution. We suggest that due to its simplicity, the
method may even be employed in noisy environments such as free-space channels.
For example, existing lidar technologies based on time--of--flight measurements
may be enhanced by the new method. On the microscale, focused classically
entangled modes may provide a new approach to precision measurements, for
example of Brownian motion in the ballistic regime in two dimensions
\cite{kheifets-observation-2014}.

\noindent\rule[1pt]{\linewidth}{1pt}
\vfill

%
%
\begin{minipage}{\textwidth}
\raisebox{0.6in}{\def\svgwidth{0.17\textwidth}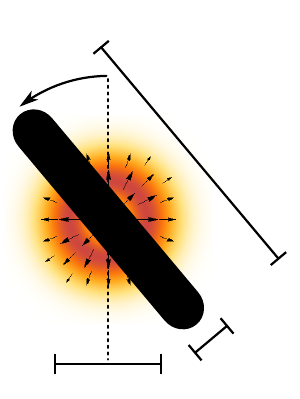}
{\def\svgwidth{0.75\textwidth}\input{fig_rotor.pgf}}
\\
\vspace{1.0cm}
\raisebox{1.65in}{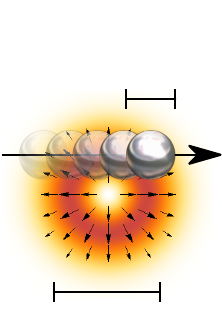}
\hspace{-1.0cm}
\input{fig_stokes.pgf}
\quad
\raisebox{0.00cm}{\input{fig_position.pgf}}
\captionof{figure}{%
    \textbf{Rotation sensing and position tracking.}
    \textbf{a.}~A metal rotor 
        (width $m = \SI{0.79(1)}{\milli\metre}$)
        turns about the centre of a radially polarized beam 
        (width $w_1 = \SI{1.95(10)}{\milli\metre}$).
        Due to the beam's classically entangled structure, the rotation in
        space induces a sinusoidal oscillation of the beam's Stokes parameters. 
        Measurements of the $s_0$, $s_1$ and $s_2$ Stokes parameters allow the
        instantaneous angle of rotation to be inferred. Each data point was
        obtained by integrating over \SI{200}{\nano\second}, so that
        electronic noise is averaged out to within the data point width. 
        Dotted curves show the theoretically expected values under the
        assumption of an ideal mode function.
    \textbf{b.}~\textit{(left)} A metal sphere 
        (diameter $d = \SI{1.00(1)}{\milli\metre}$)
        traverses a radially polarized light beam
        (width $w_2 = \SI{2.84(10)}{\milli\metre}$).
        \textit{(center)} The Stokes parameters $s_0$, $s_1$ and $s_2$ vary as
        a function of the sphere's position. 
        Solid lines show the expected Stokes parameters as obtained from
        simulation.
        \textit{(right)} The sphere's trajectory is inferred from
        the measured Stokes parameters. The sphere is moved gradually in
        discrete steps of \SI{50}{\micro\metre}, providing a calibrated,
        reproducible reference motion. To allow for a realistic comparison with
        a fast object, the acquisition time at each point is only
        \SI{250}{\pico\second}. The blue contours show the combined
        Bayesian \SI{68}{\percent} credible region $R$, while the gray shadow
        shows the sphere's dimensions to scale.
        The theoretical model used to obtain the theory and simulation curves
        for a and b, respectively, is detailed in the ``Methods'' section.
        The position tracking algorithm is described in detail in the
        Supplementary Material. 
        In both plots, $s_0$ is normalised to its initial value, while $s_1$
        and $s_2$ are normalised to the instantaneous value of $s_0$. 
}
\label{fig:results}
\end{minipage}
%
%

\clearpage
\clearpage
%
%
\begin{minipage}{\textwidth}
\shortstack{
\quad
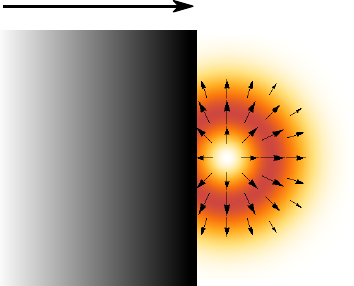 
\\ \, \\ \, \\ \, \\ \,
\\ \, \\ \, \\ \, \\ \,
{\def\svgwidth{0.15\textwidth} 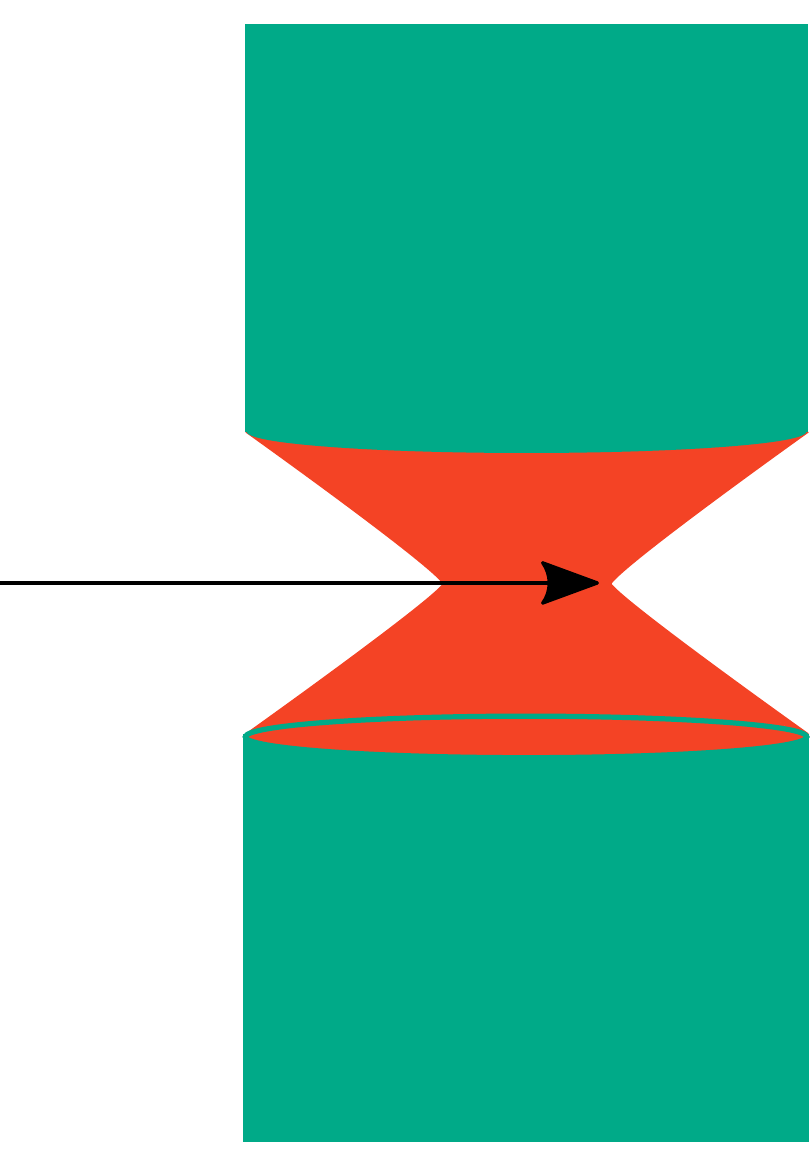}
\\ \, \\ \, \\ \, \\ \,
\\ \, \\ \, \\ \, \\ \,
}
\quad
\includegraphics[width=0.82\textwidth]{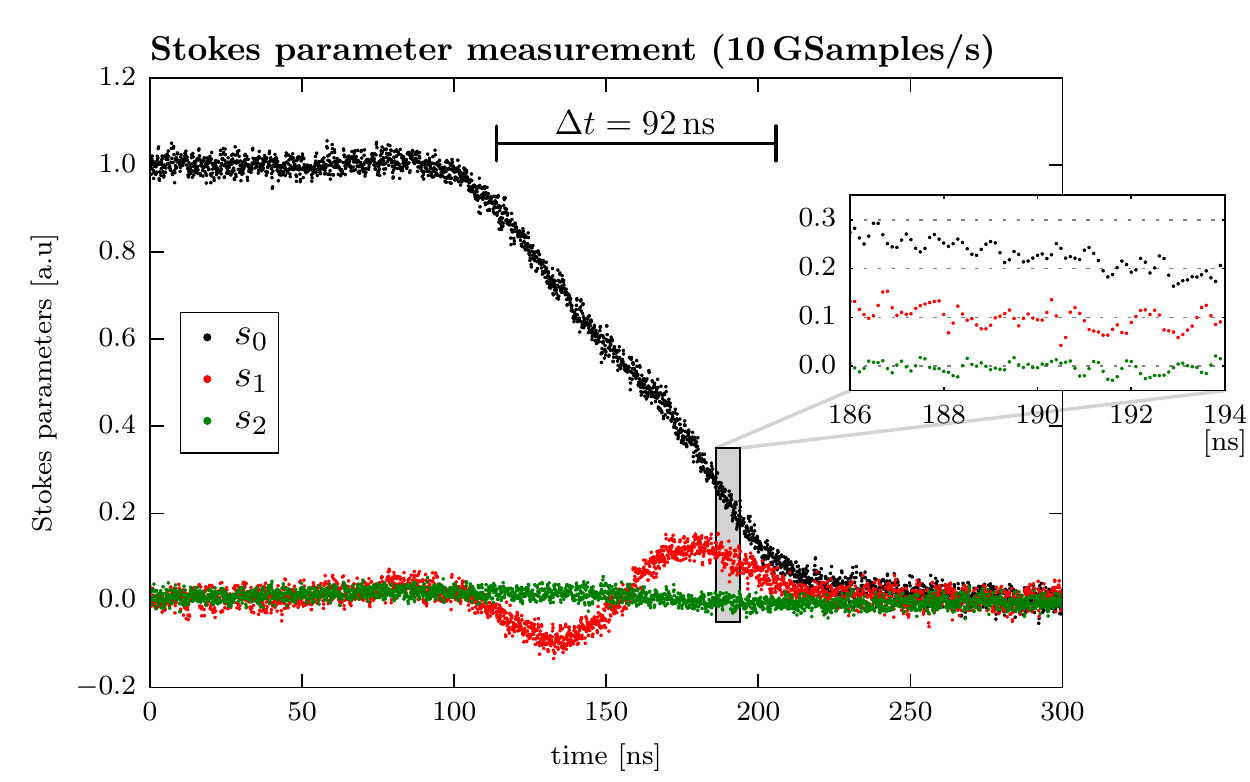}
\captionof{figure}{%
    \textbf{Real-time sensing.}
    A metal knife edge of thickness \SI{3(2)}{\micro\meter}
    cuts across a focused radially polarized mode 
    (theoretically estimated width $w_3 = \SI{2.0(5)}{\micro\meter}$) at 
    \SI{27(2)}{\meter\per\second}.
    The plot shows a sequence of single-shot measurements of the beam's $s_0$,
    $s_1$ and $s_2$ Stokes parameters during the knife edge's passage
    until the beam is fully covered (normalised to the initial power), with
    a total duration of \SI{92}{\nano\second}. 
    The sampling resolution reaches up to \SI{100}{\pico\second}. From the
    measured traces, the knife edge's direction of motion can be easily
    inferred up to a 180$^\circ$ rotation.
}
\label{fig:focusing}
\end{minipage}
%
%

\clearpage
\clearpage

\normalsize
\noindent\textbf{Methods}
\footnotesize
\smallskip

\noindent\textbf{Experimental setup.} We employed a continuous-wave diode laser
(Sumitomo SLT5411) with wavelength $\lambda=\SI{1550}{\nano\meter}$,
amplified by an erbium-doped fiber amplifier (OPREL OFA20-1231S) to a
power of \SI{50}{\milli\watt}. The initially Gaussian beam profile was
converted to a radially polarized beam using a commercially available liquid
crystal device (Arcoptix S.A.) with custom anti-reflection coatings.
The converted beam was Fourier filtered with a \SI{75}{\micro\meter} pinhole
placed between a pair of $f=\SI{50}{\milli\meter}$ plano-convex lenses and
passed two tilted waveplates in order to correct residual circular polarization
components. 
The rotor was a Cu-coated wire of width \SI{0.79(1)}{\milli\metre}, placed
in a rotation mount which was actuated by a stepper motor.
The sphere had diameter \SI{1.00(1)}{\milli\metre} and was made from steel.
It was attached to an uncoated glass substrate of thickness \SI{1}{\milli\metre}
by a meniscus of epoxy resin which was small enough to be neglected
optically.
The substrate was mounted in a vertical position on a linear precision stage
(PI M-410.DG, minimum incremental motion \SI{0.1}{\micro\metre}) and moved in
steps of \SI{50}{\micro\metre}. 
The plane of object interaction was located at \SI{50}{\centi\metre} after the
mode converter. The distances between object-camera and object-detectors were
\SI{10}{\centi\metre} and \SI{80}{\centi\metre}, respectively.  The beam had a
width of \SI{2.84(10)}{\milli\metre} in the plane of object interaction,
and a Rayleigh range of approximately \SI{2}{\metre}.
All beam widths given here and in the main text are 90-10 knife-edge widths.
The detectors used to measure the horizontal, vertical, diagonal and
anti-diagonal polarization components were two Soliton ET-3500, an Alphalas
UPD-35 and an Alphalas UPD-70, respectively. 
The detectors were temporally matched to within \SI{1.3}{\nano\second}, with the
remaining offset being compensated by the oscilloscope up to the measurement
bandwidth.
For the rotor measurement only, a pair of BPDV2120R fiber-coupled
balanced photoreceivers with \SI{43}{\giga\hertz} \SI{3}{\decibel}-bandwidth
from the company u$^2$t were used. Tailored phase plates
were used to convert the the linearly polarized Hermite-Gauss projection
modes into approximate fundamental modes in order to fiber-couple them to these
detectors.
For the real-time measurement, the beam was focused using a pair of aspheric
lenses with NA~0.65. While it is known that radially polarized beams display
a strong longitudinal electric field component on axis when focused tightly
\cite{dorn-sharper-2003}, such components are not yet predominant in the
moderate focusing regime found here. Comparative measurements with an
azimuthally polarized beam, for which no longitudinal field components appear
during focusing, showed very similar overall behavior and duration.
All signals were recorded with a digital storage oscilloscope (Agilent DSO81204B
in half-channel mode, \SI{8}{\giga\hertz} analog bandwidth,
sinc-$x$-interpolation disabled).
An InGaAs camera with resolution $320\times256$ px (Xenics XS) was used for
beam inspection and for generating spatially resolved beam polarization
tomographies prior to the actual sensing measurement. The Bayesian tracking
algorithm described in the Supplementary Material used these tomographies in
order to improve its estimate for the object position.
\\

\noindent\textbf{Theoretical model.}
In order to obtain the theoretical predictions and simulated values shown in
Figure~\ref{fig:results} we developed the following simple model.
Consider a radially polarized beam of
light of angular frequency $\omega$, propagating along the  $z$-axis of
a Cartesian reference frame and polarized in the $xy$-plane. In
the paraxial approximation its electric field takes the form
$\boldsymbol{\mathcal{E}}(\br,t) = \Re\left[\bE(\bm{\rho},z)\exp(-i \omega t) \right]$
with time-independent mode function $\bE(\bm{\rho},z)$ (Jones vector). In the
ideal case of a perfect radially polarized beam the mode function is given by
\begin{align}
\label{rad_pol_mode}
\bE(\bm{\rho},z)
=
\frac{1}{\sqrt{2}}\left[
    \hat{\bm{x}} \, \psi_{10}\!(\bm{\rho},z) + \hat{\bm{y}} \, \psi_{01}\!(\bm{\rho},z)
\right].
\end{align}
Therein $\psi_{mn}\!(\bm{\rho},z)$, with $m,n \in \{0,1,2,\ldots\}$, is the
Hermite-Gaussian solution of the paraxial wave equation of order $N=m+n$, while 
$\bm{\rho} = \hat{\bm{x}} x + \hat{\bm{y}} y$
denotes the transverse position vector. 
Please note that Eq.~(\ref{rad_pol_mode}) is a particular case of the
general expression in Eq.~(\ref{radial}).

When an opaque obstacle is placed into the beam, covering a region
$\mathcal{A}$ in the $xy$-plane, the electric field directly behind
the obstacle is given by 
\begin{align}
\label{rad_pol_mode_disturbed}
\bE'(\bm{\rho},z)=[1-\mathcal{I}_\mathcal{A}(\bm{\rho})]\bE(\bm{\rho},z), 
\end{align}
where 
\begin{align}
\label{projector}
\mathcal{I}_\mathcal{A}(\bm{\rho}) = 
\begin{cases} 
1 &\text{if } \bm{\rho} \in \mathcal{A}, \\ 0 &\text{if } \bm{\rho} \notin \mathcal{A}, 
\end{cases}
\end{align}
denotes the transmission function of the aperture complementary to the
obstruction (Babinet's principle).
 
We are interested in the perturbed beam's intensity after passing a linear
polarizer making an angle $\varphi\in\{0^\circ,90^\circ,45^\circ,135^\circ\}$
with the $x$-axis, as the Stokes parameters $s_0, s_1, s_2$ are easily
calculated from these intensities. Representing
Eqs.~(\ref{rad_pol_mode}) and (\ref{rad_pol_mode_disturbed}) as two-component
Jones vectors 
and describing the action of the linear polarizer by the projection matrix
\begin{align*}
P_\varphi=
\begin{bmatrix}
 \cos^2\varphi & \sin\varphi\cos\varphi \\
 \sin\varphi\cos\varphi & \sin^2\varphi
\end{bmatrix}
\end{align*}
yields
\begin{align}
\label{intensity_disturbed}
I_\mathcal{A}(\varphi)&=\int \left|P_\varphi \bE'(\bm{\rho},z)\right|^2\dd^2 \bm{\rho} \nonumber \\
&=\int [1-\mathcal{I}_\mathcal{A}(\bm{\rho})]\left|P_\varphi \bE(\bm{\rho},z)\right|^2\dd^2 \bm{\rho}
\end{align}
for the beam's intensity behind the polarizer. With the help of
\eref{intensity_disturbed}, the Stokes parameters, measured in the
plane of the obstructing object, take the form
\begin{subequations}
\label{stokes_parameters_disturbed}
\begin{align}
s_0 (\mathcal{A}) &= I_\mathcal{A}(0^\circ)+I_\mathcal{A}(90^\circ),\\
s_1 (\mathcal{A}) &= I_\mathcal{A}(0^\circ)-I_\mathcal{A}(90^\circ),\\
s_2 (\mathcal{A}) &= I_\mathcal{A}(45^\circ)-I_\mathcal{A}(135^\circ),
\end{align}
\end{subequations}
and are functions of $\mathcal{A}$, the region covered by the object within the
beam. In simple cases, these equations may allow for straightforward algebraic
inversion. To obtain the theoretical predictions shown in
Figure~\ref{fig:results}a (rotor), we have used the time-independent mode
function of the ideal radially polarized
beam from Eq.~(\ref{rad_pol_mode}). The rotation angle can then be obtained
analytically as $\theta_0 = \arcsin(s_1/s_2)$. The simulated reference in
Figure~\ref{fig:results}b (tracking) relies on experimentally measured
distributions of 
$\left|P_\varphi \bE(\bm{\rho},z)\right|^2$ 
that have been substituted into Eq.~(\ref{intensity_disturbed}). This allows
for the numerical evaluation of a likelihood function $L$ which is used by the
position tracking algorithm (see Supplementary Material).
\\

\normalsize
\noindent\textbf{Acknowledgements}
\footnotesize
\smallskip

\noindent The authors would like to thank Tobias R\"othlingsh\"ofer for help
    with the experiment, Irina Harder for fabricating the phase plates, and
    Thomas Bauer for useful discussions.
%

\clearpage

\noindent
\bibliographystyle{plain}

\clearpage

\section*{Supplementary information}

\textbf{Position tracking algorithm.} Here we explain the algorithm used to
infer the transverse central coordinates $(x_0,y_0)$ of a spherical object from
measurements of the polarization Stokes
parameters $\bm{s} = (s_0, s_1, s_2)$. Among the advantages of a numerical
implementation is that experimental beam imperfections and electronic detector
noise can be accounted for in a straightforward manner.
Eqs.~\eqref{stokes_parameters_disturbed} in the ``Methods'' showed how
to compute the individual Stokes parameters as a function of the
transverse position of an object covering a known region $\mathcal{A}$. Before
the actual sensing measurement, we perform this computation for each possible
object position in discrete steps using the experimentally determined spatially
resolved beam polarization tomographies
$\left|P_\varphi \bE(\bm{\rho},z)\right|^2$,
effectively yielding a look-up table $(x_0,y_0) \rightarrow \bm{s}$
as visualised by the surfaces in Supplementary Figure~\ref{fig:suppl}a. By taking into account
the normally distributed, experimentally determined statistical noise of each
detector, the look-up table can be turned into a likelihood distribution
$L\left(\bm{s}|(x_0,y_0)\right)$. During the real-time sensing measurement,
given a measured Stokes vector $\bm{s}$, the position can be inferred by
finding the coordinates $(x_0,y_0)$ maximizing this
likelihood distribution (see Supplementary Figure~\ref{fig:suppl}b).

The rotational symmetry of the radially polarized mode implies a twofold
ambiguity in determining the object's position, since rotating any object by
\SI{180}{\degree} about the beam's center leaves the Stokes parameters
invariant. This ambiguity can be lifted by making a number of additional
assumptions, such as specifying the half-plane where the particle is located at
time $t=0$ (initial position), and that its velocity is small compared to the
sampling frequency (continuity of position), or that its acceleration stays
within physically motivated bounds (continuity of velocity). Such assumptions
can easily be accommodated within the framework of Bayesian probability theory
(BPT) 
[von Toussaint, U. \emph{Rev.~Mod.~Phys.} \textbf{83,} 943--999 (2011)]
where they take the role of \textit{prior information}. BPT also assigns
a formal role to other assumptions such as the size of the object to be
tracked, and provides the means to estimate such parameters when they are not
known \textit{a priori}.  
However, in our laboratory setting we observe that the beam's symmetry is
already broken by small beam imperfections (including a slight, but not
complete, four-fold symmetry in intensity). Interestingly, we found that taking
into account this overall asymmetry helps to determine the trajectory
unambiguously with high certainty without such additional assumptions as 
described before.
If necessary, the symmetry could also be broken in more controlled
ways, e.g. by introducing a second beam, or by using a beam with a less
symmetric polarization structure 
[Beckley, A.~M. \etal \emph{Opt.~Express} \textbf{18,} 10777--10785 (2010)].

We emphasize that the algorithm used is non-recursive and does not require any
repeated iterations on the same data. Rather, it relies mainly on data which
can be pre-computed in the form of look-up tables to any desired accuracy in
a computationally efficient way. Hence, the algorithm is highly parallelizable
and thus compatible with the requirements of high-speed real-time signal
processing.  The algorithm's performance is rather insensitive to the assumed 
shape of the object: assuming a square of the same area instead of a circle
leads to identical inferred positions (up to the width of the 68\% credibility
region). Changing the assumed object size generally leads to a radial shift of
the inferred trajectory, but leaves the main features of the trajectory intact.

For the measurements shown here, we accounted for diffraction by assuming an
effective sphere diameter at the detector plane which is 35\% larger than the
sphere's actual physical dimensions.  It is likely that the accuracy can be
further increased by incorporating a more detailed model of diffraction. 
However, the actual impact of diffraction and the appropriate model to
incorporate it will depend on the particular application under consideration.
\\

\clearpage

%
\begin{figure*}
    \def\svgwidth{1.0\textwidth} 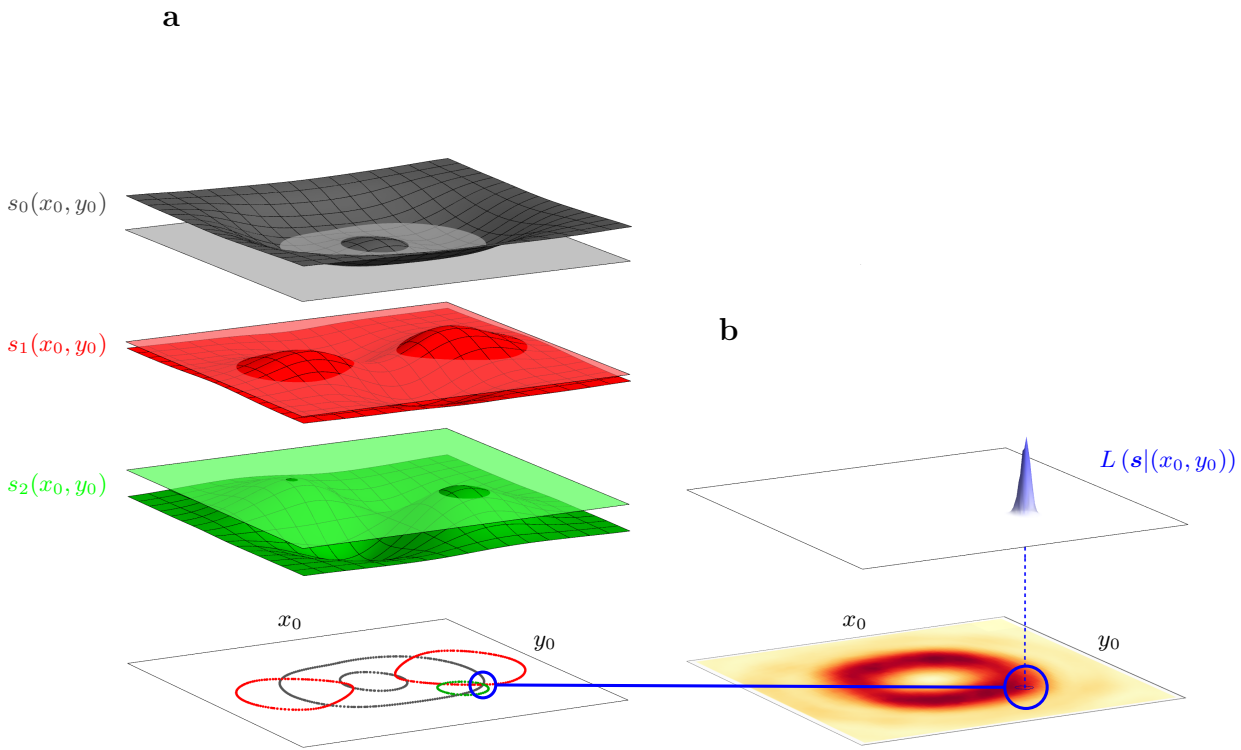
\captionof{figure}{%
    \textbf{Position tracking (experimental data).}
    \textbf{a.} For a known beam and object geometry, a look-up table
    $(x_0, y_0)~\rightarrow~s_{\{0,1,2\}}$ may be computed from
    Eqs.~\eqref{stokes_parameters_disturbed}, giving the expected Stokes
    parameters as a function of the object's transverse position within
    the beam.
    One such look-up table, obtained from experimental data, is visualized here
    as a set of three surfaces.  
    The measurement of a particular Stokes vector may be regarded as a set of
    planes intersecting these surfaces. The intersection of the
    corresponding contours (shown on the bottom) indicates the 
    position of the object.
    \textbf{b.} The upper plane shows the likelihood distribution $L(\bm{s} | (x_0,y_0))$ 
    for the measured Stokes vector $\bm{s}$. The maximum of this distribution
    corresponds to the inferred position. Note that although the theory for an
    ideal mode predicts two solutions, this symmetry may be broken in practice
    due to slight irregularities in the mode pattern, as demonstrated by this
    particular example.  
}
\label{fig:suppl}
\end{figure*}



\end{document}

%% file: fig_mode_3.pdf_tex
\begingroup%
  \makeatletter%
  \providecommand\color[2][]{%
    \errmessage{(Inkscape) Color is used for the text in Inkscape, but the package 'color.sty' is not loaded}%
    \renewcommand\color[2][]{}%
  }%
  \providecommand\transparent[1]{%
    \errmessage{(Inkscape) Transparency is used (non-zero) for the text in Inkscape, but the package 'transparent.sty' is not loaded}%
    \renewcommand\transparent[1]{}%
  }%
  \providecommand\rotatebox[2]{#2}%
  \ifx\svgwidth\undefined%
    \setlength{\unitlength}{293.96818848bp}%
    \ifx\svgscale\undefined%
      \relax%
    \else%
      \setlength{\unitlength}{\unitlength * \real{\svgscale}}%
    \fi%
  \else%
    \setlength{\unitlength}{\svgwidth}%
  \fi%
  \global\let\svgwidth\undefined%
  \global\let\svgscale\undefined%
  \makeatother%
  \begin{picture}(1,0.93682992)%
    \put(0,0){\includegraphics[width=\unitlength]{fig_mode_3.pdf}}%
    \put(0.15484124,0.0801345){\makebox(0,0)[lb]{\smash{$s_1$}}}%
    \put(0.45827169,0.17458281){\makebox(0,0)[lb]{\smash{$s_2$}}}%
    \put(0.25944338,0.41502141){\makebox(0,0)[lb]{\smash{$s_3$}}}%
    \put(0.65152492,0.07972765){\makebox(0,0)[lb]{\smash{$s_1$}}}%
    \put(0.95495537,0.17417563){\makebox(0,0)[lb]{\smash{$s_2$}}}%
    \put(0.75612705,0.41461457){\makebox(0,0)[lb]{\smash{$s_3$}}}%
    \put(0.040928,0.88362805){\color[rgb]{0,0,0}\makebox(0,0)[lb]{\smash{\bf \large a}}}%
    \put(0.54808167,0.88007486){\color[rgb]{0,0,0}\makebox(0,0)[lb]{\smash{\bf \large b}}}%
  \end{picture}%
\endgroup%

%% file: fig_setup_8.pdf_tex
\begingroup%
  \makeatletter%
  \providecommand\color[2][]{%
    \errmessage{(Inkscape) Color is used for the text in Inkscape, but the package 'color.sty' is not loaded}%
    \renewcommand\color[2][]{}%
  }%
  \providecommand\transparent[1]{%
    \errmessage{(Inkscape) Transparency is used (non-zero) for the text in Inkscape, but the package 'transparent.sty' is not loaded}%
    \renewcommand\transparent[1]{}%
  }%
  \providecommand\rotatebox[2]{#2}%
  \ifx\svgwidth\undefined%
    \setlength{\unitlength}{1008bp}%
    \ifx\svgscale\undefined%
      \relax%
    \else%
      \setlength{\unitlength}{\unitlength * \real{\svgscale}}%
    \fi%
  \else%
    \setlength{\unitlength}{\svgwidth}%
  \fi%
  \global\let\svgwidth\undefined%
  \global\let\svgscale\undefined%
  \makeatother%
  \begin{picture}(1,0.50714286)%
    \put(0,0){\includegraphics[width=\unitlength]{fig_setup_8.pdf}}%
    \put(0.505798,0.33408566){\color[rgb]{0,0,0}\makebox(0,0)[b]{\smash{\bf BS$_1$}}}%
    \put(0.59103675,0.38332224){\color[rgb]{0,0,0}\makebox(0,0)[b]{\smash{\bf BS$_2$}}}%
    \put(0.72504558,0.45474277){\color[rgb]{0,0,0}\makebox(0,0)[b]{\smash{\bf PBS}}}%
    \put(0.7260033,0.29918721){\color[rgb]{0,0,0}\makebox(0,0)[b]{\smash{\bf PBS}}}%
    \put(0.15917819,0.12295033){\color[rgb]{0,0,0}\makebox(0,0)[rb]{\smash{\bf cw laser}}}%
    \put(0.11401312,0.24055963){\color[rgb]{0,0,0}\makebox(0,0)[lt]{\begin{minipage}{0.08872528\unitlength}\raggedright \bf mode\\ converter\end{minipage}}}%
    \put(0.65631452,0.33954525){\color[rgb]{0,0,0}\makebox(0,0)[b]{\smash{$\lambda/2$}}}%
    \put(0.47965078,0.12250349){\color[rgb]{0,0,0}\makebox(0,0)[lt]{\begin{minipage}{0.24565861\unitlength}\raggedright \bf b \quad object\\ \quad\quad interaction\end{minipage}}}%
    \put(0.29752166,0.06987927){\color[rgb]{0,0,0}\makebox(0,0)[lt]{\begin{minipage}{0.25842593\unitlength}\raggedright \bf a \quad mode\\ {\footnotesize~~} \quad preparation\end{minipage}}}%
    \put(0.66177415,0.17228312){\color[rgb]{0,0,0}\makebox(0,0)[lt]{\begin{minipage}{0.23992414\unitlength}\raggedright \bf c \quad verification\\ \quad\quad camera\end{minipage}}}%
    \put(0.84389479,0.22491412){\color[rgb]{0,0,0}\makebox(0,0)[lt]{\begin{minipage}{0.26135847\unitlength}\raggedright \bf d \quad Stokes parameter\\ \quad\quad measurement\end{minipage}}}%
  \end{picture}%
\endgroup%

%% file: fig_rotor_schematic.pdf_tex
\begingroup%
  \makeatletter%
  \providecommand\color[2][]{%
    \errmessage{(Inkscape) Color is used for the text in Inkscape, but the package 'color.sty' is not loaded}%
    \renewcommand\color[2][]{}%
  }%
  \providecommand\transparent[1]{%
    \errmessage{(Inkscape) Transparency is used (non-zero) for the text in Inkscape, but the package 'transparent.sty' is not loaded}%
    \renewcommand\transparent[1]{}%
  }%
  \providecommand\rotatebox[2]{#2}%
  \ifx\svgwidth\undefined%
    \setlength{\unitlength}{86.27230072bp}%
    \ifx\svgscale\undefined%
      \relax%
    \else%
      \setlength{\unitlength}{\unitlength * \real{\svgscale}}%
    \fi%
  \else%
    \setlength{\unitlength}{\svgwidth}%
  \fi%
  \global\let\svgwidth\undefined%
  \global\let\svgscale\undefined%
  \makeatother%
  \begin{picture}(1,1.32889666)%
    \put(0,0){\includegraphics[width=\unitlength]{fig_rotor_schematic.pdf}}%
    \put(0.22611546,0.93543182){\color[rgb]{0,0,0}\makebox(0,0)[lb]{\smash{$\theta$}}}%
    \put(0.04986428,1.27488889){\color[rgb]{0,0,0}\makebox(0,0)[lb]{\smash{\textbf{\large a}}}}%
    \put(0.36107319,0.01474256){\color[rgb]{0,0,0}\makebox(0,0)[b]{\smash{$w_1$}}}%
    \put(0.7288949,0.0882324){\color[rgb]{0,0,0}\makebox(0,0)[lb]{\smash{$m$}}}%
    \put(0.52969588,1.03098003){\color[rgb]{0,0,0}\makebox(0,0)[lb]{\smash{$l \gg w_1$}}}%
  \end{picture}%
\endgroup%

%% file: fig_tracking_schematic.pdf_tex
\begingroup%
  \makeatletter%
  \providecommand\color[2][]{%
    \errmessage{(Inkscape) Color is used for the text in Inkscape, but the package 'color.sty' is not loaded}%
    \renewcommand\color[2][]{}%
  }%
  \providecommand\transparent[1]{%
    \errmessage{(Inkscape) Transparency is used (non-zero) for the text in Inkscape, but the package 'transparent.sty' is not loaded}%
    \renewcommand\transparent[1]{}%
  }%
  \providecommand\rotatebox[2]{#2}%
  \ifx\svgwidth\undefined%
    \setlength{\unitlength}{63.79558773bp}%
    \ifx\svgscale\undefined%
      \relax%
    \else%
      \setlength{\unitlength}{\unitlength * \real{\svgscale}}%
    \fi%
  \else%
    \setlength{\unitlength}{\svgwidth}%
  \fi%
  \global\let\svgwidth\undefined%
  \global\let\svgscale\undefined%
  \makeatother%
  \begin{picture}(1,1.44159544)%
    \put(0,0){\includegraphics[width=\unitlength]{fig_tracking_schematic.pdf}}%
    \put(0.00668698,1.36855944){\color[rgb]{0,0,0}\makebox(0,0)[lb]{\smash{\textbf{\large b}}}}%
    \put(0.48359052,0.01993672){\color[rgb]{0,0,0}\makebox(0,0)[b]{\smash{$w_2$}}}%
    \put(0.68118862,1.06718995){\color[rgb]{0,0,0}\makebox(0,0)[b]{\smash{$d$}}}%
  \end{picture}%
\endgroup%

%% file: fig_chopper_schematic.pdf_tex
\begingroup%
  \makeatletter%
  \providecommand\color[2][]{%
    \errmessage{(Inkscape) Color is used for the text in Inkscape, but the package 'color.sty' is not loaded}%
    \renewcommand\color[2][]{}%
  }%
  \providecommand\transparent[1]{%
    \errmessage{(Inkscape) Transparency is used (non-zero) for the text in Inkscape, but the package 'transparent.sty' is not loaded}%
    \renewcommand\transparent[1]{}%
  }%
  \providecommand\rotatebox[2]{#2}%
  \ifx\svgwidth\undefined%
    \setlength{\unitlength}{101.78787537bp}%
    \ifx\svgscale\undefined%
      \relax%
    \else%
      \setlength{\unitlength}{\unitlength * \real{\svgscale}}%
    \fi%
  \else%
    \setlength{\unitlength}{\svgwidth}%
  \fi%
  \global\let\svgwidth\undefined%
  \global\let\svgscale\undefined%
  \makeatother%
  \begin{picture}(1,0.80827266)%
    \put(0,0){\includegraphics[width=\unitlength]{fig_chopper_schematic.pdf}}%
  \end{picture}%
\endgroup%

%% file: fig_focusing_schematic.pdf_tex
\begingroup%
  \makeatletter%
  \providecommand\color[2][]{%
    \errmessage{(Inkscape) Color is used for the text in Inkscape, but the package 'color.sty' is not loaded}%
    \renewcommand\color[2][]{}%
  }%
  \providecommand\transparent[1]{%
    \errmessage{(Inkscape) Transparency is used (non-zero) for the text in Inkscape, but the package 'transparent.sty' is not loaded}%
    \renewcommand\transparent[1]{}%
  }%
  \providecommand\rotatebox[2]{#2}%
  \ifx\svgwidth\undefined%
    \setlength{\unitlength}{233.0992595bp}%
    \ifx\svgscale\undefined%
      \relax%
    \else%
      \setlength{\unitlength}{\unitlength * \real{\svgscale}}%
    \fi%
  \else%
    \setlength{\unitlength}{\svgwidth}%
  \fi%
  \global\let\svgwidth\undefined%
  \global\let\svgscale\undefined%
  \makeatother%
  \begin{picture}(1,1.44128532)%
    \put(0,0){\includegraphics[width=\unitlength,page=1]{fig_focusing_schematic.pdf}}%
    \put(0.6451496,1.12230275){\color[rgb]{1,1,1}\makebox(0,0)[b]{\smash{\textbf{{NA} 0.65}}}}%
    \put(0.6451496,0.22139896){\color[rgb]{1,1,1}\makebox(0,0)[b]{\smash{\textbf{{NA} 0.65}}}}%
    \put(-0.00000001,0.75316757){\color[rgb]{0,0,0}\makebox(0,0)[lb]{\smash{knife edge}}}%
    \put(0,0){\includegraphics[width=\unitlength,page=2]{fig_focusing_schematic.pdf}}%
    \put(0.47249491,1.51413141){\color[rgb]{0,0,0}\makebox(0,0)[lt]{\begin{minipage}{0.35795073\unitlength}\raggedright \end{minipage}}}%
  \end{picture}%
\endgroup%

%% file: fig_suppl_300dpi.pdf_tex
\begingroup%
  \makeatletter%
  \providecommand\color[2][]{%
    \errmessage{(Inkscape) Color is used for the text in Inkscape, but the package 'color.sty' is not loaded}%
    \renewcommand\color[2][]{}%
  }%
  \providecommand\transparent[1]{%
    \errmessage{(Inkscape) Transparency is used (non-zero) for the text in Inkscape, but the package 'transparent.sty' is not loaded}%
    \renewcommand\transparent[1]{}%
  }%
  \providecommand\rotatebox[2]{#2}%
  \ifx\svgwidth\undefined%
    \setlength{\unitlength}{704.93347361bp}%
    \ifx\svgscale\undefined%
      \relax%
    \else%
      \setlength{\unitlength}{\unitlength * \real{\svgscale}}%
    \fi%
  \else%
    \setlength{\unitlength}{\svgwidth}%
  \fi%
  \global\let\svgwidth\undefined%
  \global\let\svgscale\undefined%
  \makeatother%
  \begin{picture}(1,0.55083937)%
    \put(0,0){\includegraphics[width=\unitlength,page=1]{fig_suppl_300dpi.pdf}}%
    \put(0.52767046,0.31117416){\color[rgb]{0,0,0}\makebox(0,0)[lb]{\smash{\large \bf b}}}%
    \put(0.11507443,0.54449093){\color[rgb]{0,0,0}\makebox(0,0)[lb]{\smash{\large \bf a}}}%
    \put(0.20073153,0.09888748){\color[rgb]{0,0,0}\makebox(0,0)[lb]{\smash{$x_0$}}}%
    \put(0.39019747,0.08389511){\color[rgb]{0,0,0}\makebox(0,0)[lb]{\smash{$y_0$}}}%
    \put(0.61947393,0.09888748){\color[rgb]{0,0,0}\makebox(0,0)[lb]{\smash{$x_0$}}}%
    \put(0.80893998,0.08389511){\color[rgb]{0,0,0}\makebox(0,0)[lb]{\smash{$y_0$}}}%
    \put(0,0.40798445){\color[rgb]{0,0,0}\makebox(0,0)[lb]{\smash{\textcolor{darkgray}{$s_0 (x_0,y_0)$}}}}%
    \put(0,0.3033439){\color[rgb]{0,0,0}\makebox(0,0)[lb]{\smash{\textcolor{red}{$s_1 (x_0,y_0)$}}}}%
    \put(0,0.19870335){\color[rgb]{0,0,0}\makebox(0,0)[lb]{\smash{\textcolor{green}{$s_2 (x_0,y_0)$}}}}%
    \put(0.81019771,0.21692195){\color[rgb]{0,0,0}\makebox(0,0)[lb]{\smash{\textcolor{blue}{$L \left(\bm{s} | (x_0,y_0)\right) $}}}}%
  \end{picture}%
\endgroup%